\documentclass[aps,nofootinbib,groupedaddress]{revtex4}
\usepackage{graphicx,amsmath,amssymb}

\newcommand{\be}{\begin{eqnarray}}
\newcommand{\ee}{\end{eqnarray}}

\begin{document}
\author{P. Faccioli$^1$\footnote{Talk given at the 2nd International 
Conference on Nuclear and Particle Physics at Jefferson Laboratory, 
Dubrovnik, May 2003.}, A. Schwenk$^2$ and E.V. Shuryak$^3$}
\email{faccioli@ect.it, aschwenk@mps.ohio-state.edu, 
shuryak@tonic.physics.sunysb.edu}
\affiliation{$^1$E.C.T.*, Strada delle Tabarelle 286, 
Villazzano (Trento), I-38050, Italy\\
$^2$Department of Physics, The Ohio State University, 
Columbus, OH 43210, USA\\
$^3$Department of Physics and Astronomy, State University 
of New York, Stony Brook, NY 11794-3800, USA}

\title{Instanton contribution to the pion and proton 
electro-magnetic formfactors at $Q^2 \gtrsim 1 \, \mathrm{GeV}^2$}

\begin{abstract}
Studying the instanton-induced contributions to various hard 
exclusive reactions provides physical insight into the transition 
from the non-perturbative to the perturbative regime of QCD. To this end, 
we compute the leading-instanton contributions to the electro-magnetic
and transition formfactors using an effective theory of the 
instanton liquid model. We report predictions for the electro-magnetic
formfactor $F_\pi(Q^2)$ of the pion as well as novel results for the 
proton Dirac formfactor $F_1(Q^2)$.
\end{abstract}

\maketitle

\section{Introduction}

The transition from the non-perturbative to the perturbative 
regime of QCD is fundamental to our understanding of the strong 
interactions. For this purpose, the study of exclusive electro-magnetic 
reactions at intermediate and high momenta plays a prominent role. As
for structure functions from inclusive deep inelastic processes, elastic 
and transition formfactors encode valuable information about the 
short-distance structure of the hadrons. In contrast, however, exclusive 
reactions are sensitive to the non-perturbative forces responsible for 
the recombination of the scattered partons into the final state.

\begin{figure}[t!]
\includegraphics[scale=0.33,clip=]{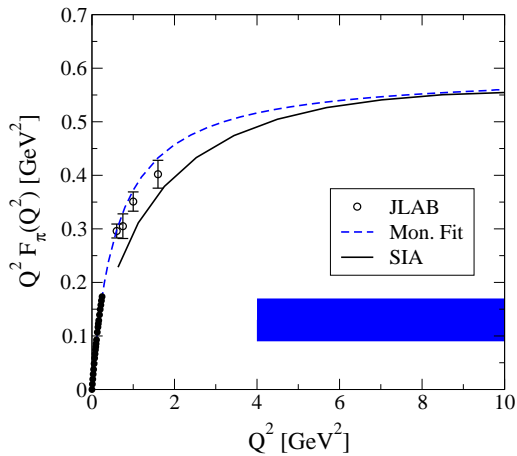}
\caption{The recent JLAB data for $Q^2 \, F_\pi(Q^2)$~\cite{jlab} 
in comparison with the asymptotic PQCD prediction (thick bar, for a 
typical $\alpha_s \approx 0.2-0.4$), the monopole fit (dashed line), 
and the results of the SIA prediction (solid line). The SIA
calculation is not reliable below $Q^2 \sim 1 \, \text{GeV}^2$.
The solid circles at low $Q^2$ denote the SLAC data~\cite{SLAC}.}
\label{data1}
\end{figure}

It is now becoming clear that the asymptotic perturbative regime is 
not yet reached in the electro-magnetic formfactors, even at surprisingly 
large momentum transfers. This conclusion follows form the results of 
two high-precision experiments performed at Jefferson Laboratory (JLAB), 
namely the measurement of the formfactor of the charged pion and of the
ratio of electric to magnetic formfactors $G_E(Q^2)/G_M(Q^2)$ for the proton.
The pion formfactor has been measured very accurately for momentum 
transfers $0.6 \, \text{GeV}^2 < Q^2 < 1.6 \, \text{GeV}^2$ by the 
$F_\pi$ collaboration~\cite{jlab}. Not only are the data at highest 
experimentally accessible momenta still very far from the asymptotic 
limit, but also the trend is away from the perturbative QCD (PQCD) prediction 
(see Fig.~\ref{data1}). This observation contrasts the result of 
the CLEO experiment for the $\gamma \gamma^\star \pi^0$ neutral pion 
transition formfactor, where the asymptotic PQCD regime is reached 
already for $Q^2 \sim 2 \, \text{GeV}^2$~\cite{CLEO}.

The ratio $\mu \, G_E(Q^2)/G_M(Q^2)$ for the proton ($\mu$ is the magnetic 
moment) has been obtained from recoil polarization measurements up to 
$Q^2 < 5.6 \, \text{GeV}^2$~\cite{JLAB1,JLAB2}. At low momenta, this 
ratio approaches 1, supporting the older SLAC results, 
which lead to the conclusion that the proton electric 
and magnetic formfactors can be very well described by the same 
dipole fit, $G^p_{E(M)\,dip} = e(\mu)/(1+Q^2/M_{dip}^2)^2$ with
$M_{dip}=0.84~\textrm{GeV}$.  However, at larger momenta, $Q^2 \gtrsim 2\, 
\text{GeV}^2$, the electric formfactor falls off faster than the magnetic
one, and the ratio $\mu \, G_E(Q^2)/G_M(Q^2)$ decreases significantly,
in disagreement however with results based on the Rosenbluth separation 
method.

The slope of the ratio $\mu \, G_E(Q^2)/G_M(Q^2)$ versus $Q^2$ is sizable and
indicates that the asymptotic prediction is not supported by the available 
experimental data. In fact, the naive $Q^2$ power counting predicts
the same scaling in $Q^2$ for both electric and magnetic formfactors.

It is natural to ask what dynamics is responsible for the 
deviation from the asymptotic behavior.
By comparing the formfactor of the charged pion to the neutral pion 
transition formfactor, one concludes that strong non-perturbative 
dynamics is at work in the former reaction and that it is much weaker 
in the transition formfactor. Such dynamics must be responsible for
delaying the onset of the perturbative regime in the formfactor of the
charged pion.

An important question to address is to what extent the observed behavior 
of the pion and proton formfactors can be understood in terms of the  
non-perturbative dynamics associated with the spontaneous breaking of 
chiral symmetry. Clearly, one expects that these forces should play a
prominent role in the pion formfactor, due to its Goldstone boson nature.
They similarly influence the proton formfactors. This is most evident 
in the small momentum regime, where a dynamically generated quark mass 
provides a source for quark helicity flip, therefore contributing to 
the Pauli formfactor $F_2(Q^2)=(G_M(Q^2)-G_E(Q^2))/(1+Q^2/(4M^2))$.

Instantons are non-perturbative gauge configurations which have been shown
to play an important role in the dynamical breaking of chiral 
symmetry (\cite{dyakonovchiral,shuryakrev} for a recent study see also
\cite{scalar}), as they naturally 
generate a density of quasi-zero modes of the Dirac operator. These gluon 
fields are related to the tunneling between degenerate vacua of 
QCD. In order to account for their contribution 
to the correlation functions in QCD, one needs to know how often a 
tunneling event occurs (i.e., the typical instanton density, $\bar{n}$) 
and how long it lasts for (i.e., the typical instanton size, $\bar{\rho}$). 
In the Instanton Liquid Model (ILM), these parameters are extracted 
phenomenologically from the global properties of the QCD 
vacuum~\cite{shuryak82}. This leads to $\bar{n} \simeq 1 \, \text{fm}^{-4}$ 
and $\bar{\rho} \simeq 1/3 \, \text{fm}$.

The instanton contribution to the pion and nucleon formfactors has been 
the subject of several studies over the last few years. Early analyses, 
however, were either model-dependent or could only make indirect
contact with experiments. In~\cite{forkel}, Forkel and Nielsen computed
the pion formfactor in a sum-rule approach, which takes into account 
the instanton contribution in the operator product expansion.
As in sum-rule approaches, this calculation required a phenomenological 
description of the continuum of excitations. In order to avoid this 
model dependence, the relevant electro-magnetic pion and proton three-point 
functions were calculated in coordinate space in~\cite{blotz,3ptILM}. 
The result can then be compared to phenomenological estimates of the 
same correlation functions, obtained by Fourier transforming fits of 
the experimental data. In this way, the unknown contribution from the 
continuum of excitations could be excluded by considering sufficiently 
large-sized correlation functions. Unfortunately, this method has the 
shortcoming that it does not allow a direct comparison of the 
theoretical predictions in coordinate space to the experimental data.

A direct comparison between ILM predictions and experiments became 
possible by combining the Single-Instanton-Approximation (SIA) developed
in~\cite{SIA,thesis} with the mixed time-momentum representation widely
used in lattice calculations~\cite{draper89}. This lead to predictions for the 
pion and nucleon formfactors~\cite{pionFF,nucleonFF}, and the pion
and nucleon correlators, with particular attention to their energy 
dispersions~\cite{mymasses}. The SIA is an effective theory of the ILM, 
in which the degrees of freedom of the instanton closest to the propagating
quarks are treated explicitly, while the contribution of all other 
pseudo-particles in the vacuum is encoded in a single parameter,
the quark effective mass~\cite{SIA}. The combined framework has the 
advantage that one can compute correlation functions in momentum space, 
which include the dominant short-distance effects. Moreover, this approach
does not have the above discussed model dependences. Clearly, in the SIA 
it is possible to compute accurately only correlation functions which 
are dominated by single-instanton effects. This, for example, prevents 
one from making predictions in the small-momentum regime, typically 
below $1 \, \text{GeV}$ (for a detailed discussion see~\cite{mymasses}).

\section{Pion and nucleon formfactors in the SIA}

The details of the calculation of the formfactors in the SIA
are given in~\cite{pionFF,nucleonFF}. The formfactors are obtained 
from appropriate ratios of three- to two-point correlation functions. 
For example, the pion formfactor is given by
\be
\label{FF}
\frac{G^{(3)}_4(t,{\bf q}/2; -t,- {\bf q}/2)}{G^{(2)}(2 t,{\bf q}/2)} \to 
F_\pi(Q^2) ,
\ee
where $G^{(3)}_4$ denotes the pion three-point correlator with fourth 
component of the electro-magnetic current. In the time-momentum 
representation, the three-point function $G^{(3)}_\mu$ is given by
\be
\label{3p}
G^{(3)}_\mu(t,{\bf p}+{\bf q};-t,{\bf p}) = \int d^3{\bf x} \,
d^3{\bf y} \, e^{-i \, {\bf p} \cdot {\bf x} + i \,
({\bf p}+{\bf q}) \cdot {\bf y}}
\langle 0 | \, j_5(t,{\bf y}) \, J_\mu(0,{\bf 0}) \,
j_5^\dagger(-t,{\bf x}) \, | 0 \rangle .
\ee
In the ratio for the formfactor, Eq.~(\ref{FF}), 
the propagation of a pion which is struck by a
virtual photon is normalized by the propagation in the absence of the
external probe. To this end, one needs the pion two-point correlator 
$G^{(2)}$, which is defined analogously
\be
\label{2p}
G^{(2)}(2 t,{\bf p}) = \int d^3{\bf x} \,
e^{i \, {\bf p} \cdot {\bf x}} \, \langle 0 | \,
j_5(t,{\bf x}) \, j_5^\dagger(-t,{\bf 0}) \, | 0 \rangle .
\ee  
Here, the pseudo-scalar current $j_5(x)=\bar{u}(x) \, i\gamma_5 \,
d(x)$ excites states with the quantum numbers of the pion and
$J_\mu(0)$ denotes the electro-magnetic current operator.
Similar expressions lead to the nucleon formfactors~\cite{nucleonFF}. 
We also note that this method does not require the wave function of the 
hadron as input.

Let us first discuss the results obtained for the pion 
formfactor~\cite{pionFF}. The SIA prediction is shown in 
Fig.~\ref{data1} in comparison to the JLAB data and the monopole 
fit. For momenta where the SIA is a valid approximation, we observe 
that our results are consistent with the available experimental data, 
and coincide with the monopole form in the kinematic region 
accessible to JLAB. This result complements the analysis of Blotz 
and Shuryak~\cite{blotz}, where a similar agreement with the
monopole fit was found at small momentum transfers. Clearly, 
upcoming measurements at JLAB will be able to test the single-instanton 
prediction as the microscopic non-perturbative mechanism at intermediate 
momentum transfers.

We find that the single-instanton contribution remains well above 
the PQCD prediction, throughout the region under present and upcoming 
investigations at JLAB. This nicely contrasts the situation 
for the $\gamma \gamma^\star \pi^0$ transition formfactor, 
where the asymptotic PQCD regime is reached already for $Q^2 \sim
2 \, \text{GeV}^2$~\cite{CLEO}. This striking difference is explained 
by instanton arguments as well, since the three-point function 
in this case has a different chiral structure. As a result, the 
instanton contribution to the $\gamma \gamma^\star \pi^0$ reaction
is suppressed with respect to the 
corresponding contribution to the formfactor of the charged  pion. 
Physically, this 
is the same reason why the vector and axial channels have a rather 
strong ``Zweig'' rule, forbidding flavor mixing, while for the 
pseudo-scalars such a mixing is very strong.

Next, we consider the results for the formfactors of the proton.
In~\cite{nucleonFF}, the electric formfactor was computed
in the same approach. We found that the SIA prediction roughly 
followed the dipole fit. In order to extend the calculation to the 
magnetic formfactor, one needs to identify a suitable three-point 
function which relates directly to the physical formfactor and
also receives a maximally enhanced single-instanton contribution.
The correlation function satisfying these requirements is given by
\be
G^{(3)\,M}_2(t,{\bf q}/2;-t,-{\bf q}/2) = \int d^3{\bf x} \,
d^3{\bf y} \, e^{i \, {\bf q} \cdot ({\bf x}+{\bf y})/2}
\langle 0 | \, \textrm{Tr} [ \, \eta_{sc}(t,{\bf y}) \, J_2(0,{\bf 0}) \,
\bar{\eta}_{sc}(-t,{\bf x}) \,\gamma_2 \, ] \, | 0 \rangle,
\ee
where ${\bf q}$ is chosen along the $\widehat{\bf 1}$ direction and
$\eta_{sc}(x) = \epsilon^{a b c} \, [ u_a^T(x) C \gamma_5 d_b(x) ]
\, u_c(x)$ is the nucleon scalar current, which excites states with the 
quantum numbers of the proton ($C$ is the charge conjugation matrix).

Although the SIA leads to reasonable results for both formfactors, 
it turns out that $G_E(Q^2)$  and $G_M(Q^2)$ receive rather large 
contributions from many-instanton effects, even for relatively high 
momenta, $Q^2 \sim 1-4 \, \text{GeV}^2$, where one would naively expect 
the single-instanton contribution to be dominant. However, by analyzing 
the SIA contribution to the relevant three-point correlators, 
we have found that the single-instanton contribution to the Dirac
formfactor $F_1(Q^2)=(G_E(Q^2)+Q^2/(4 M^2)\, G_M(Q^2))/(1 + Q^2/(4 M^2))$ is enhanced
and gives a realistic $F_1(Q^2)$ already for $Q^2 \gtrsim 1 \, 
\text{GeV}^2$, while the SIA contribution to the Pauli formfactor 
$F_2(Q^2)$ is suppressed. As a consequence, in the SIA one should 
only expect to obtain a realistic prediction for the electric and 
magnetic formfactors in the kinematic region where the Dirac formfactor 
dominates over the Pauli formfactor.

\begin{figure}[t!]
\includegraphics[scale=0.33,clip=]{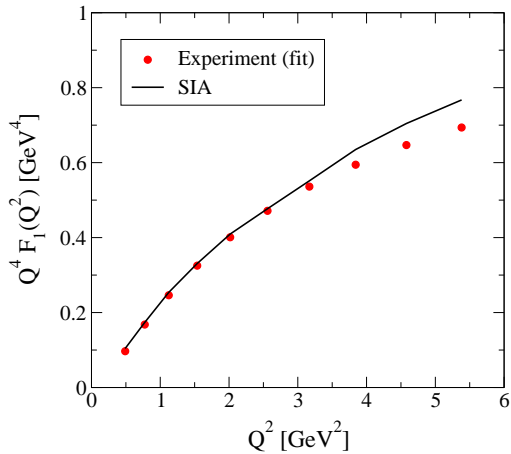}
\caption{
The SIA prediction for the proton Dirac formfactor (solid line)
compared to a fit of the experimental data (circles). 
The fit has been obtained by assuming a traditional dipole fit for the 
magnetic 
formfactor, $G_M(Q^2)=\mu/(1+Q^2/0.71)^2$, and then extracting $G_E(Q^2)$
from the JLAB parametrization of the electric over magnetic formfactor ratio: 
$\mu\,G_E(Q^2)/G_M(Q^2)=1-0.13(Q^2-0.04)$, with $Q$ in $\text{GeV}$. 
Asymptotic PQCD gives $Q^4\,F_1(Q^2) \sim \text{const}$~\cite{brodsky}.}
\label{F1}
\end{figure}

The SIA prediction for $F_1(Q^2)$ is presented in Fig.~\ref{F1}. These are
preliminary novel results. The details of this calculation will be reported 
in a separate publication~\cite{F1pub}. The agreement between the 
theoretical calculation and the extracted experimental data is
striking. This implies that the 't~Hooft interaction is able to
account for the important non-perturbative dynamics in the proton.

Finally, we qualitatively discuss the instanton contributions to the 
Pauli formfactor $F_2(Q^2)$ of the proton, which receives large contributions
from two- and more instantons and thus cannot be addressed in the SIA. 
The instanton contributions to the Pauli formfactor arise from these
collective many-instanton effects, which scale with the square root of 
the instanton diluteness. The dominant of such instanton-induced collective
phenomena is the dynamical breaking of chiral symmetry, which gives rise
to a momentum-dependent effective quark mass. Moreover, it has been 
recently shown by Kochelev that quarks in the instanton vacuum acquire an 
intrinsic Pauli formfactor~\cite{kochelev}. It was observed that the 
size of the constituent quarks and the sign of their formfactor agree 
with the values extracted phenomenologically by Petronzio {\it et
al.}~\cite{simula1}. Both quark effective masses and the Pauli formfactor 
of the constituent quarks enhance the helicity-flipping transitions, 
and therefore contribute to $F_2(Q^2)$. A realistic assessment of the 
contributions of many-instanton effects to the Pauli formfactor 
of the proton requires challenging numerical simulations in the ILM,
which are currently being performed.

\section{Conclusions}

We have reported our recent results for the instanton contribution
to the pion and proton electro-magnetic formfactors. These 
are the lightest meson and baryon in the spectrum, where the 
instanton-induced effects are largest.
The SIA makes it possible to perform the calculations for the 
formfactors directly in momentum space. Our results depend only on two 
phenomenological parameters of the ILM, which are fixed from global
properties of the QCD vacuum.\footnote{Note that the SIA prediction of 
the pion formfactor does not depend on the instanton density, but only 
on the instanton size, which is extracted from lattice simulations}
The SIA results agree with experiment, where the corresponding correlators 
receive contributions from a single instanton. This is the case for
the electro-magnetic formfactor of the pion and the Dirac formfactor of the
proton. Moreover, the SIA makes predictions for the kinematic 
regime under investigation at JLAB and constrains the onset of the 
asymptotic PQCD regime to relatively high momenta for the pion
formfactor. Our results imply that instanton-induced forces are able to
account for the electro-magnetic structure of the light hadrons 
at short distances.

Studying the instanton-induced contribution to the hadron formfactors 
is also important from a broader standpoint, in order to understand 
the transition from the non-perturbative to the perturbative regime of 
QCD. We have shown that the 't~Hooft interaction provides an 
explanation why the perturbative regime is reached much later in the
pion formfactor than for the $\gamma \gamma^* \pi_0$ transition 
formfactor. In order to study the effect of instantons on the
electro-magnetic formfactor of the proton in the same transition 
window as for the pion
formfactor, one has to evaluate the Pauli formfactor, $F_2(Q^2)$,
in the ILM. This problem is currently under numerical 
investigation.


\begin{thebibliography}{99}
%
%
\bibitem{jlab}
J. Volmer {\it et al.}, Phys. Rev. Lett. \textbf{86} (2001) 1713.
\bibitem{CLEO}
J. Gronberg {\it et al.}, Phys. Rev. \textbf{D57} (1998) 33.
\bibitem{SLAC}
S.R. Amendolia {\it et al.}, Nucl. Phys. \textbf{B277} (1986) 168.
%
%
\bibitem{JLAB1} M.K. Jones, \emph{et al.}, Phys. Rev. Lett. 
\textbf{84} (2000) 1398.
\bibitem{JLAB2} O. Gayou, \emph{et al.}, Phys. Rev. Lett.
\textbf{88} (2002) 092301.
%
%
\bibitem{dyakonovchiral} D. Diakonov, Chiral Symmetry
Breaking by Instantons, Lectures given at the Enrico Fermi
School in Physics, Varenna, 1995, hep-ph/9602375.
\bibitem{shuryakrev}
T. Sch\"afer and E.V. Shuryak, Rev. Mod. Phys. \textbf{70} (1998) 323.
\bibitem{scalar} 
P. Faccioli and T.A. DeGrand, Phys. Rev. Lett. (2003) in press, 
hep-ph/0304219.
\bibitem{shuryak82}
E.V. Shuryak, Nucl. Phys. \textbf{B214} (1982) 237.
\bibitem{forkel}
H. Forkel and M. Nielsen, Phys. Lett. \textbf{B345} (1997) 55.
\bibitem{blotz}
A. Blotz and E.V. Shuryak, Phys. Rev. \textbf{D55} (1997) 4055.
\bibitem{3ptILM}
P. Faccioli and E.V. Shuryak, Phys. Rev. \textbf{D65} (2002)
076002.  
\bibitem{SIA} P. Faccioli and E.V. Shuryak, Phys. Rev. 
\textbf{D64} (2001) 114020.
\bibitem{thesis} P. Faccioli, Ph.D. Thesis, SUNY Stony Brook, 2002, 
unpublished. 
\bibitem{draper89}
T. Draper, R.M. Woloshyn, W. Wilcox and K.F. Liu, Nucl. Phys. \textbf{B318}
(1989) 319.
\bibitem{pionFF} P. Faccioli, A. Schwenk and E.V. Shuryak, 
Phys. Rev. \textbf{D67} (2003) 113009.
\bibitem{nucleonFF} P. Faccioli, A. Schwenk and E.V. Shuryak, Phys. Lett.
\textbf{B549} (2002) 93.
\bibitem{mymasses} P. Faccioli, Phys. Rev. \textbf{D65} (2002) 094014.
\bibitem{F1pub} P. Faccioli and E.V. Shuryak, in preparation.
\bibitem{kochelev} N.I. Kochelev, Phys. Lett. \textbf{B565} (2003) 131. 
\bibitem{simula1} R. Petronzio, S. Simula and G. Ricco, Phys. Rev.
\textbf{D67} (2003) 094009.
\bibitem{brodsky} S.J. Brodsky and G.R. Ferrar, Phys. Rev. Lett. 
\textbf{31} (1973) 1153.\\
G.P. Lepage and S.J. Brodsky, Phys. Rev. Lett. \textbf{43} (1979) 545. 
\end{thebibliography}
\end{document}